\documentclass{PoS}

\usepackage{amsmath,amssymb}
\usepackage{txfonts}
\usepackage{textcomp}
\usepackage{fontenc,times,mathptmx,graphicx}
\usepackage{comment}

\newcommand{\be}{\begin{equation}}
\newcommand{\ee}{\end{equation}}
\newcommand{\beq}{\begin{equation}}
\newcommand{\eeq}{\end{equation}}
\newcommand{\bea}{\begin{eqnarray}}
\newcommand{\eea}{\end{eqnarray}}

\newcommand{\alphaem}{\alpha_{\rm em}}
\newcommand{\alphas}{\alpha_{\rm s}}
\newcommand{\RIMOMprime}{{RI\textquotesingle -MOM}}
\newcommand{\RIprime}{{RI\textquotesingle }}
\newcommand{\QCD}{{\rm QCD}}
\newcommand{\QED}{{\rm QED}}
\newcommand{\aep}{\frac{\alphaem}{4\pi}}

\newcommand{\pslash}{{\not{\hspace{-0.08cm}p}}}

\title{Non-perturbative renormalization in QCD+QED and its application to weak decays}

\ShortTitle{Non-perturbative renormalization in QCD+QED}

\author{\speaker{Matteo Di Carlo}\,\thanks{{matteo.dicarlo@roma1.infn.it}}~,~Guido Martinelli\\
        Dipartimento di Fisica, Universit\`a di Roma La Sapienza and INFN, Sezione di Roma\\
        Piazzale Aldo Moro 5, 00185 Rome, Italy
     }

\author{Davide Giusti~,~Vittorio Lubicz\\
        Dipartimento di Matematica e Fisica, Universit\`a Roma Tre and INFN, Sezione di Roma Tre\\
        Via della Vasca Navale 84, 00146 Rome, Italy
  }
  
\author{Christopher T. Sachrajda\\
        Department of Physics and Astronomy, University of Southampton\\
        Southampton SO17 1BJ, UK
  }  
  
  \author{Francesco Sanfilippo~,~Silvano Simula\\
        Istituto Nazionale di Fisica Nucleare, Sezione di Roma Tre\\
        Via della Vasca Navale 84, 00146 Rome, Italy
  }

\author{Nazario Tantalo\\
        Dipartimento di Fisica and INFN, Universit\`a di Roma ``Tor Vergata'' \\
        Via della Ricerca Scientifica 1, 00133 Rome, Italy
  }

\abstract{We present a novel strategy to  renormalize lattice operators in QCD+QED, including first order QED corrections to the non-perturbative evaluation of QCD renormalization constants. Our procedure takes systematically into account the mixed non-factorizable QCD+QED effects which were neglected in previous calculations, thus significantly reducing the systematic uncertainty on renormalization corrections. The procedure is presented here in the \RIMOMprime~scheme, but it can be applied to other schemes (e.g. RI-SMOM) with appropriate changes. We  discuss the application of this strategy to the calculation of the leading isospin breaking corrections to the leptonic decay rates $\Gamma(\pi_{\mu 2})$ and $\Gamma(K_{\mu 2})$, evaluated for the first time on the lattice. The precision in the matching to the $W$-regularization scheme is improved to $\mathcal{O}(\alphaem\alphas(M_W))$ with respect to previous calculations. Finally, we show the updated precise result obtained for the Cabibbo-Kobayashi-Maskawa matrix element $|V_{us}|$.  }

\FullConference{37th International Symposium on Lattice Field Theory - Lattice2019\\
		16-22 June 2019\\
		Wuhan, China}

\begin{document}

\section{Introduction}
Current lattice  calculations are generally performed in the isospin symmetric limit of QCD, in which the up and down quark masses are taken to be equal ($m_u=m_d$) and electromagnetic (e.m.) interactions are switched off. These have reached a precision below the percent level for many hadronic quantities~\cite{Aoki:2019cca}. This implies that further improvements in the  determination of physical observables useful to extract the Cabibbo-Kobayashi-Maskawa (CKM) matrix elements and  test the limits of the Standard Model are only possible if strong and e.m. isospin breaking (IB) effects, which are expected to be of  $\mathcal{O}(1\%)$, are included in lattice calculations. A possible way to include such effects in lattice simulations is to use a perturbative approach (the \emph{RM123 method} proposed in Refs.~\cite{deDivitiis:2011eh,deDivitiis:2013xla}) in which the lattice path-integral is expanded in terms of the two small parameters $\alphaem$ and $(m_d-m_u)/\Lambda_\QCD$ and  IB corrections to observables computed in the iso-symmetric limit are evaluated at first order in these parameters. Such a perturbative approach also allows one to control the subtraction of infrared (IR) divergences arising when evaluating  QED corrections to  hadron decay rates. While IR divergences are cancelled by including both virtual corrections and the real emission of photons, ultraviolet (UV) divergences have to be treated by including QED corrections in the renormalization procedure. When e.m. corrections at $\mathcal{O}(\alphaem)$ are added to QCD, renormalization constants (RCs) of composite operators can be written as 
\beq
\label{eq:Z_expand}
Z=\left( 1 + \aep\,  \Delta Z \right) \, Z^\QCD~,
\eeq
where $Z^\QCD$ is the RC computed in pure QCD and $\Delta Z$ represents the correction introduced by e.m. interactions. In general $\Delta Z$ and $Z^{\QCD}$ are matrices which mix different operators, and hence the order in which they are multiplied defines the correction $\Delta Z$.
 The calculation of  $\Delta Z$ has been so far performed in the so-called \emph{factorization approximation}, in which the correction to the RC is simply evaluated as $\Delta Z\equiv \Delta Z^\QED$, 
 namely the pure e.m. correction at $\mathcal{O}(\alphaem)$  
  computed in perturbation theory through the evaluation of one-loop diagrams.
  In this approximation  mixed non-factorizable QCD+QED contributions to the RCs are 
  neglected and hence a systematic uncertainty is introduced in the calculation. In this talk we present a novel framework to compute non-perturbatively such mixed contributions, thus overcoming the factorization approximation and improving the precision of RCs. In the following, we will focus on the renormalization of the operator $O_1$ entering the effective Hamiltonian which describes the leptonic decay of a  pseudoscalar meson,
\beq
 \label{eq:Heff}
 \mathcal{H}_{\rm eff} = \frac{G_F}{\sqrt{2}} \, V^*_{q_1 q_2}\, O_1 \equiv \frac{G_F}{\sqrt{2}} \, V^*_{q_1 q_2}\, [ \bar{q}_2 \gamma_\mu (1-\gamma_5) q_1 ] [ \bar{\nu}_\ell \gamma^\mu (1-\gamma_5) \ell ]~,
 \eeq 
 but the  discussion can also be applied to other lattice operators (e.g. quark bilinear operators). A method  to compute the decay rate of such processes on the lattice has been proposed in Ref.~\cite{Carrasco:2015xwa} and, as a first step, the leading IB corrections to the ratio 
 $\Gamma[K\to\mu\nu_\mu(\gamma)]/\Gamma[\pi\to\mu\nu_\mu(\gamma)]$ have been computed~\cite{Giusti:2017dwk}. In the ratio there is a large cancellation of renormalization corrections and the factorization approximation only affects the quark mass RCs. However, to obtain a separate determination of the first order corrections to the kaon and pion decay rates it is necessary to overcome the factorization approximation and renormalize $O_1$ non-perturbatively on the lattice. In the following we discuss how to implement the non-perturbative renormalization in QCD+QED and show its impact on the evaluation of leptonic decay rates of kaons and pions. All the details of the calculation of the decay rates 
  are given in Ref.~\cite{DiCarlo:2019thl}, while a specific paper on the QCD+QED renormalization is in preparation \cite{DiCarlo:ren}.

\section{Leading IB corrections to light meson leptonic decay rates}
When studying QED corrections  at order $\mathcal{O}(\alphaem)$ to the leptonic decay of a pseudoscalar meson $P$, diagrams with both virtual and real photons, which are separately IR divergent, must be considered in order to get a finite result for the decay rate~\cite{Bloch:1937pw},
\beq
\Gamma(P_{\ell2}) \equiv \Gamma_P^{(0)} \, (1+\delta R_P) = \Gamma_0(P\to\ell\nu_\ell) + \Gamma_1 (P\to\ell\nu_\ell \gamma)~,
\eeq
where $\Gamma_P^{(0)}$ is the 
 decay rate in pure QCD, $\delta R_P$  its correction at  first order in $\alphaem$ and $(m_d-m_u)/\Lambda_\QCD$ and the subscript {\footnotesize{${0,1}$}} 
 denotes the number of photons in the final state. The initial proposal of Ref.~\cite{Carrasco:2015xwa} was to consider sufficiently soft photons, emitted in the meson rest frame with a maximum energy $\Delta E_\gamma$, such that they do not resolve the internal structure of the meson. The meson can then be treated as a point-like particle and $\Gamma_1 (P\to\ell\nu_\ell \gamma)\simeq\Gamma_1(\Delta E_\gamma)$ can be computed in perturbation theory. 
The non-perturbative evaluation of $\Gamma_1$ is now in progress as reported at this conference~\cite{Kane:2019jtj,deDivitiis:2019uzm}. In this framework, we are interested in the calculation of the correction $\delta R_P$ for light-mesons. This quantity gets two kinds of contributions: one coming from the strong IB and e.m. corrections to the 
 amplitude 
 and one coming from QED corrections to the renormalization of  $O_1$, namely
 { \beq
\label{eq:deltaRP}
\delta R_P=\delta R_P^{\,  ampl} + \delta R^{\, ren}~.
\eeq}\noindent
The first term has been already computed  in Ref.~\cite{Giusti:2017dwk} for kaons and pions in the evaluation of the first order correction to the ratio $\Gamma(K_{\mu2})/\Gamma(\pi_{\mu2})$, while the calculation of the second one has been addressed in Ref.~\cite{DiCarlo:2019thl} and is the aim of this talk\footnote{Since the correction $\delta R^{\, ren}$  only depends on the operator mediating the process, it follows that $\delta R_K^{ren}=\delta R_\pi^{ren}$ and therefore its contribution cancels out in the calculation of $\delta R_{K\pi}=\delta R_K - \delta R_\pi$ done in Ref.~\cite{Giusti:2017dwk}.}.

\section{Non-perturbative renormalization in QCD+QED}
\begin{figure}[h]
\centering
     \includegraphics[width=0.85\textwidth]{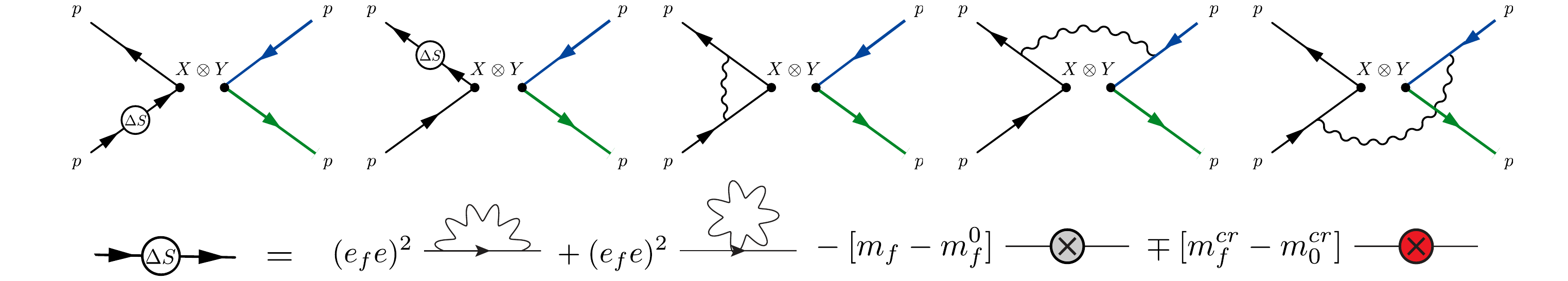}
     \caption{Diagrams considered to compute the corrections to the Green function of the operators $O_{1-5}$ (\emph{first line}) and to the quark propagator (\emph{second line}). The last two diagrams in the second line represent the mass and critical Wilson parameter counter-terms.}
     \label{fig:diags}
\end{figure}

The calculation of e.m. effects must be consistent at $\mathcal{O}(\alphaem)$ with the value of $G_F$ extracted from the lifetime of the muon. The most natural way of proceeding is to renormalize $O_1$ in the $W$-regularization scheme~\cite{Sirlin:1981ie} (see Sec.~II of Ref.~\cite{Carrasco:2015xwa} for details),
in which the effective Hamiltonian of Eq.~\eqref{eq:Heff} gets a finite correction and the operator $O_1^\textrm{W-reg}$ is renormalized using a properly regularized photon propagator. Since we are not able to implement the $W$-regularization directly in lattice calculations, the inverse lattice spacing being much smaller than the $W$-boson mass $M_W$, the calculation takes place in two steps. We start by renormalizing the four-fermion operator $O_1$ on the lattice in a given scheme, e.g. in the \RIMOMprime~scheme \cite{Martinelli:1994ty}, non-perturbatively in QCD and at $\mathcal{O}(\alphaem)$ in QED, taking into account possible mixing with other lattice operators. 
Then, the renormalized operator $O_1^\textrm{\RIprime}(\mu)$ is perturbatively matched to the corresponding operator renormalized in the $W$-regularization, 
\beq
O_1^\textrm{W-reg}(M_W) = Z^\textrm{W-\RIprime} \left(M_W/\mu, \alphas(\mu), \alphaem\right) \, O_1^\textrm{\RIprime}(\mu).
\eeq
The coefficient $Z^\textrm{W-\RIprime}$ can be computed in perturbation theory first by evolving the operator in the \RIprime~scheme from the scale $\mu$ to $M_W$ and then by matching it to the corresponding operator in the $W$-scheme.
 By including  the two-loop anomalous dimension of  $\mathcal{O}(\alphaem\alphas)$ in the evolution operator, the residual truncation error of the matching is of $\mathcal{O}(\alphaem\alphas(M_W))$, reduced from  $\mathcal{O}(\alphaem\alphas(1/a))$ of Ref.~\cite{Carrasco:2015xwa}.
We choose to renormalize the operator $O_1$ in the \RIMOMprime~scheme, but the same procedure can be  applied to other schemes such as RI-SMOM with the appropriate modifications of kinematics and projectors \cite{Sturm:2009kb}. The use of twisted-mass fermions implies that the operator $O_1$ mixes with  four other lattice operators with different chiralities\footnote{\scriptsize
$
O_{1,2}^{\rm bare}=[ \bar{q}_2 \gamma_\mu (1\mp\gamma_5) q_1 ] [ \bar{\nu}_\ell \gamma^\mu (1-\gamma_5) \ell ]
$,  $
O_{3,4}^{\rm bare}=[ \bar{q}_2  (1\mp\gamma_5) q_1 ] [ \bar{\nu}_\ell  (1+\gamma_5) \ell ]
$, $
O_{5}^{\rm bare}=[ \bar{q}_2  \sigma_{\mu\nu}(1+\gamma_5) q_1 ] [ \bar{\nu}_\ell  \sigma^{\mu\nu}(1+\gamma_5) \ell ]
$.
}, due to the explicit breaking of chiral symmetry,
\beq
\label{eq:ORI_Obare}
O_1^\text{\RIprime}(\mu) = \sum_{k=1}^{5} \left[ Z_O(\mu a) \right]_{1k}\, O_k^{\rm bare}(a)~.
\eeq  
The matrix $Z_O$ can be obtained by applying the \RIMOMprime~condition on $\Gamma_O$, the matrix obtained projecting the amputated 
Green functions on the tree-level Dirac structures of the operators,
 \beq 
Z_O(\mu a) \, \prod_f \left(Z_f(\mu a)\right)^{-1/2}\, \Gamma_O(pa) \big|_{p^2=\mu^2} = \hat{1}~,
\eeq 
where $ Z_O$ and $Z_f^{-1/2}$ are respectively the RCs of the operator and of  fermion fields, the latter being defined in the \RIMOMprime~scheme as
 \beq 
Z_f(\mu a) = - \frac{i}{12} {\rm Tr} \left[ \frac{\pslash \,  \langle S_f(pa)\rangle^{-1}}{p^2}\right]\Bigg|_{p^2=\mu^2}~.
 \eeq 
Expanding the RCs in terms of $\alphaem$ as in Eq.~\eqref{eq:Z_expand}, we obtain the  condition for the correction $\Delta Z_O$,
\beq
\label{eq:DeltaZO}
\Delta Z_O(\mu a)= - \prod_f \left( Z_f^\QCD(\mu a)\right)^{-1/2} \, Z_O^\QCD(\mu a)  \,  \Delta \Gamma_O(\mu a) + \frac 12 \sum_f \Delta Z_f(\mu a) ~ .
\eeq
To summarize, 
$\Delta Z_O$ is a combination of the corrections to the  fermion field RCs,  the corrections to  the projected Green function  $\Gamma_O$ and the pure QCD RCs. 
This applies also to $\Delta Z_O^\QED$, where all the ingredients must be evaluated in the absence of QCD, namely
\beq
\label{eq:DeltaZO_QED}
\Delta Z_O^\QED(\mu a)= - \Delta \Gamma_O^\QED(\mu a) + \frac 12 \sum_f \Delta Z_f^\QED(\mu a) \, .
\eeq
The diagrams necessary for the computation of $\Delta \Gamma_O$  and $\Delta Z_f$ in the electro-quenched approximation are depicted in Fig.~\ref{fig:diags} and they have been evaluated on the lattice both in QCD+QED and in pure QED. Photon propagators in Fig.~\ref{fig:diags} are realized using stochastic photon fields generated from a Gaussian distribution as in Ref.~\cite{Giusti:2017dmp}, thus avoiding the usage of all-to-all quark propagators.
More details are given in Ref.~\cite{DiCarlo:ren}.
\subsection{QCD+QED renormalization beyond the factorization approximation}
We find it convenient to introduce the ratio 
\beq
\label{eq:ratio}
\mathcal{R} \equiv (Z^\QED)^{-1} Z (Z^\QCD)^{-1}  = {1} + \aep \, \eta~,
\eeq
and  decompose any RC as
 \beq
 \label{eq:Zdecomp}
Z=Z^\QED \, \mathcal{R} \, Z^\QCD = \left[ {1}+\aep (\Delta Z^\QED + \eta) \right]\,Z^\QCD~,
\eeq 
where 
 \beq
\label{eq:eta}
\eta=\Delta Z - \Delta Z^\QED
\eeq
quantifies the violation to the factorization approximation and encodes all the mixed QCD+QED contibutions not included in the product $\Delta Z^\QED Z^\QCD$. 
 The advantage of computing $\eta$ is that in the ratio \eqref{eq:ratio} the pure QCD and pure QED discretization effects and anomalous dimensions cancel out. 
Moreover, an attractive feature of $\eta$ is that if $\Delta Z^\QED$ is computed on the lattice
 using the \emph{same} stochastic photon fields as for the calculation of $\Delta Z$, the statistical uncertainty related to the photon sampling is significantly reduced in the difference~\eqref{eq:eta} for several entries of the matrix $\eta$. 

\subsection{Numerical analysis}

\begin{figure}[h]
\centering
     \includegraphics[width=0.45\textwidth]{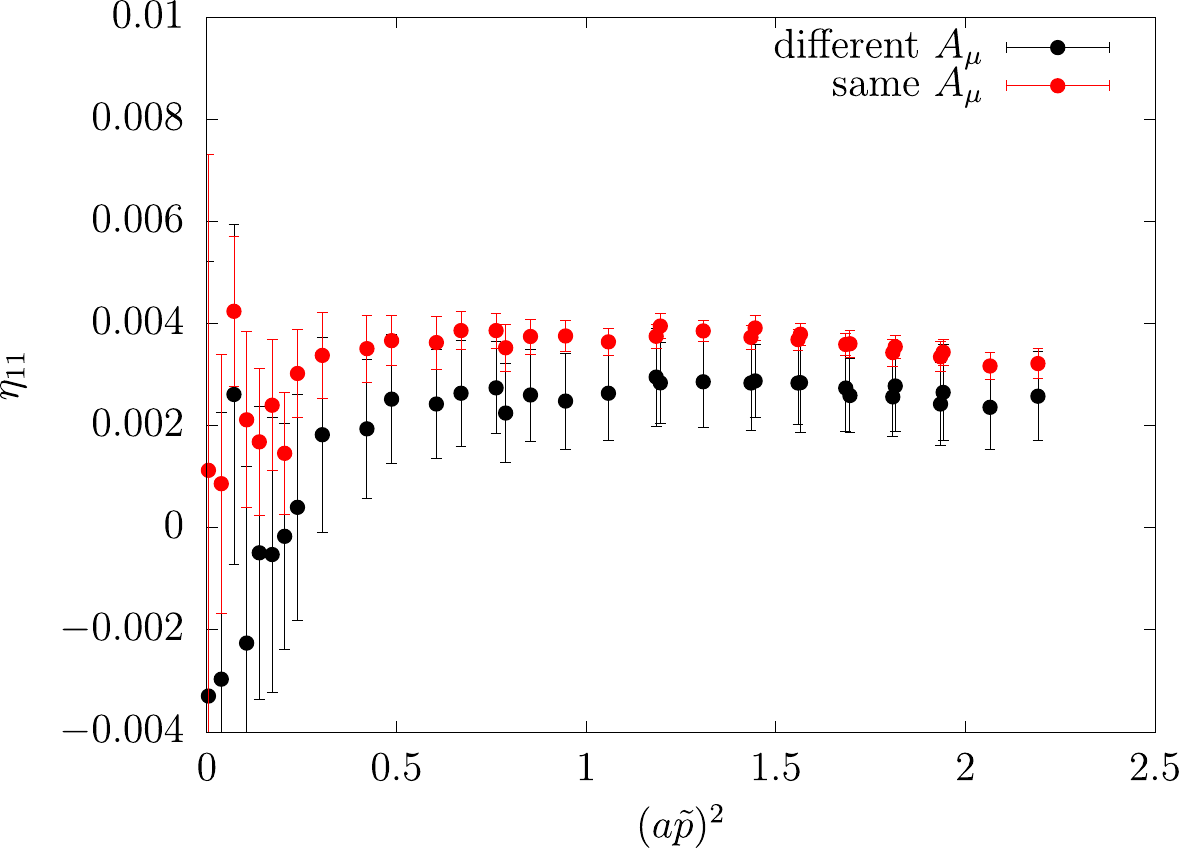}
     \includegraphics[width=0.45\textwidth]{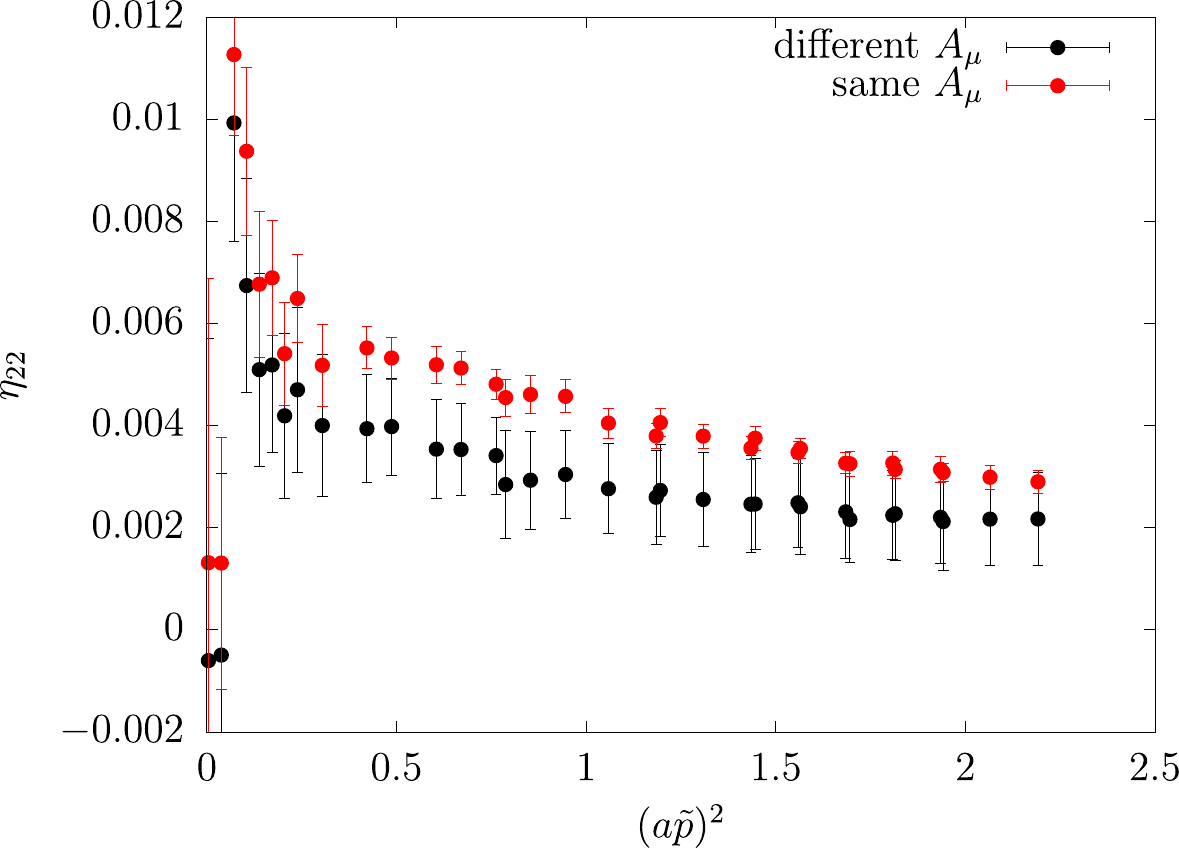}
     \caption{Results for $\eta_{11}$ and $\eta_{22}$ obtained using the same or different photon fields $A_\mu$ in the QCD+QED and the pure QED analyses. 
     The stochastic noise is about 5 times smaller when using the same fields.
     }
     \label{fig:etacorruncorr}
\end{figure}

In our analysis, we have performed two different calculations, one in QCD+QED and one in pure QED at $\mathcal{O}(\alphaem)$, with the same lattice parameters and stochastic photon fields but with the QCD links set to 1. We have used gauge configurations with $N_f=4$ degenerate dynamical quarks produced by the ETM Collaboration~\cite{Carrasco:2014cwa}.
The first step in the determination of the RCs is the computation of the Green functions in the two theories, at different values of the external momenta $(a\tilde{p})^2 = \sum_\mu \sin^2(ap_\mu)$, followed by the calculation of the corrections using Eqs.~\eqref{eq:DeltaZO} and \eqref{eq:DeltaZO_QED}. 
Since RI-MOM is a mass independent scheme, a chiral extrapolation is needed. Particular attention has to be paid when extrapolating Green functions involving scalar or pseudoscalar currents, which in pure QCD suffer from the contamination of Goldstone poles $\propto 1/M_P^2$. Indeed, when including QED corrections,  double poles of the form $ \Delta M_P^2/M_P^4$ are generated in the correction to the Green function and have to be properly subtracted.
Once the mass dependence is removed from the RCs (both in the valence and in the sea), we  compute the matrix $\eta_O$ according to Eq.~\eqref{eq:eta}. 
  In Fig.~\ref{fig:etacorruncorr} we present the results for $\eta_{11}$ and $\eta_{22}$ obtained using the same or different stochastic photon fields and we notice that the statistical uncertainty is reduced by approximately a factor of 5 in the former case for both the matrix entries, as mentioned above.
In order to remove the dependence on the renormalization scale,  $\eta$ is evolved to the reference scale $a\mu=1$ using the two-loop anomalous dimension matrix  of $\mathcal{O}(\alphas\alphaem)$~\cite{DiCarlo:ren}. 
Once the matrix $\eta_O$ is evaluated, the correction $\Delta Z_O$ is obtained as $\Delta Z_O = \eta_O + [\Delta Z_O^\QED]_{an}$, i.e. adding back the one-loop QED RC computed this time \emph{analytically} in perturbation theory.
We have finally reconstructed the mixing of the lattice operators (see Eq.~\eqref{eq:ORI_Obare}) at  $\mathcal{O}(\alphaem)$ and we are now able to derive the operator $O_1^\text{W-reg}$ renormalized in the $W$-regularization scheme by means of a perturbative matching. From the term of $\mathcal{O}(\alphaem)$ of  
$O_1^\text{W-reg}(M_W)$ it is possible to extract the term $\delta R^{\,ren}$ of Eq.~\eqref{eq:deltaRP} and the detailed calculation can be found in Ref.~\cite{DiCarlo:2019thl}. We remind the reader that  knowledge of  $\delta R_P^{\,ampl}$ alone allows one to extract only the ratio of $K$ and $\pi$ decay rates. Now, instead, we are able to determine separately the two quantities $\delta R_\pi$ and $\delta R_K$.

\section{Results}
The results obtained for the leading IB corrections to light-meson leptonic decay rates~\cite{DiCarlo:2019thl} are
\beq
\label{eq:results}
\delta R_{\pi^\pm} = 0.0153~(19)~, \quad 
\delta R_{K^\pm} = 0.0024~(10)~.
\eeq
Our results can be compared with those obtained in Chiral Perturbation Theory ($\chi$PT) and currently adopted by the PDG~\cite{PDG},  $\delta R_{\pi^\pm} = 0.0176(21)$ and $\delta R_{K^\pm} = 0.0064(24)$ respectively. The difference is within 1$\sigma$ for $\delta R_{\pi}$, but is somewhat larger for $\delta R_{K}$. Our determination of $\delta R_{K}$ has an uncertainty a factor of about 2.4 smaller than the one obtained in $\chi$PT and such improvement depends crucially on the non perturbative calculation of the mixing presented in this talk. Indeed, we can compare the results in Eq.~\eqref{eq:results} with preliminary estimates~\cite{Giusti:2018puc} computed in the factorization approximation (i.e. assuming $\eta=0$), namely
$\delta R_{\pi^\pm}^{\eta=0} = 0.0148\,(26)~$ and $\delta R_{K^\pm}^{\eta=0} = 0.0020\,(20)$, where a conservative systematic uncertainty was added to account for the missing non-factorizable contributions in the QCD+QED~renormalization. 
The improvement of the uncertainties confirms the need for a non-perturbative calculation of $\eta$.
The introduction of $\eta$ allows also to update the result of the ratio of kaon and pion decay rates, $\delta R_{K\pi}$, previously obtained in Ref.~\cite{Giusti:2017dwk}, where the factorization approximation was applied to the RCs of quark masses. The updated result is $\delta R_{K\pi} = -0.0126~(14)$. Combining these results with the experimental measurements of the decay rates and the value $|V_{ud}|=0.97420(21)$ from superallowed nuclear $\beta$ decays
~\cite{Hardy:2016vhg}, we are able to extract the CKM matrix element
\beq
\label{eq:Vus}
|V_{us}| = 0.22538~(46)~.
\eeq
Our result is in agreement with the latest estimate $|V_{us}|=0.2253(7)$~\cite{PDG}, but it improves the uncertainty by a factor of approximately $1.5$. Taking the values $|V_{ub}|=0.00413(49)$~\cite{PDG} and $|V_{ud}|=0.97420(21)$~\cite{Hardy:2016vhg}, our result in Eq.~\eqref{eq:Vus} implies that the unitarity of the first row of the CKM matrix is confirmed to better than the per-mille level,
$|V_{ud}|^2 + |V_{us}|^2 + |V_{ub}|^2 = 0.99988~(46)$.
\section{Conclusions}
We have presented a new strategy to renormalize operators non-perturbatively on the lattice with the inclusion of e.m. corrections of $\mathcal{O}(\alphaem)$ and the non-factorizable QCD+QED contributions, which had been neglected in previous calculations, have now been systematically included. By introducing the two-loop anomalous dimension at $\mathcal{O}(\alphaem\alphas)$ in the matching we have reduced the residual uncertainty to $\mathcal{O}(\alphaem\alphas(M_W))$. The renormalization procedure has been presented here in the \RIMOMprime~scheme, but it can be extended to other schemes with appropriate modifications. 
The method has been applied to the calculation of the leading IB corrections to $\Gamma(\pi_{\mu 2})$ and $\Gamma(K_{\mu 2})$, where the introduction of the non-factorizable terms in the mixing yielded a significant improvement of the precision of final results. The details of the calculation are discussed in Refs.~\cite{DiCarlo:2019thl,DiCarlo:ren}.
  We have presented the first result of $|V_{us}|$ obtained from a first-principle calculation, with a better precision than that currently quoted in the PDG Review~\cite{PDG}, and we have tested the unitarity of the first row of the CKM matrix with a relative uncertainty smaller than one per-mille.

\acknowledgments
V.L., G.M. and S.S. thank MIUR (Italy) for partial support under the
contract PRIN 2015. C.T.S. was supported by an Emeritus Fellowship from the Leverhulme Trust.
N.T. thanks the Univ. of Rome Tor Vergata for the support granted to the project PLNUGAMMA.

\end{document}